\documentclass[
    ,final            
  ]{aipproc}

\layoutstyle{6x9}

\begin{document}

\title{Light Scalar Puzzle in QCD}

\classification{PACS Nos.: 13.75.Lb, 11.15.Pg, 11.80.Et,
12.39.Fe}

\keywords      {Effective chiral Lagrangian, QCD, Light scalar mesons}

\author{Amir H. Fariborz}{
  address={Department of Mathematics/Science, State University of New York
  Institute of Technology, Utica, New York 13504-3050, USA}}

\author{Renata Jora}{
  address={Grup de Fisica Teorica, Universitat Autonoma de Barcelona,
  E-08193 Belaterra (Barcelona), Spain}}
  
\author{Joseph Schechter (speaker)}{
  address={Physics Department, Syracuse University,
  Syracuse, NY 13244-1130, USA}}

\begin{abstract}
 An approach to understanding the light scalar meson
 spectroscopy is briefly reviewed.   
\end{abstract}

\maketitle

\section{Discussion}

     The main concern at the moment in particle physics might be
     called the $3.5 \times 10^9$ dollar question (the rough cost of 
     LHC). Specifically, where is the Higgs scalar
     boson and what are its properties?
     
     The topic under discussion might be called the $3.5 \times 10^1$ cent
     question. That is roughly the cost of making a few Xerox copies
     of old published experimental data concerning the s-wave channel
     in low energy pion pion scattering. Specifically we can ask, is there a 
     light scalar meson in this channel?
     
     Are these two topics related? The Higgs sector of the standard
     model is well known to be formally identical to the SU(2) linear
     sigma model, which has often been used to describe low energy pion physics.
     In addition to the pions, the model contains the light scalar called
     sigma. Of course the scales of the two models are very different-
     about 100 MeV for the pion physics case and about 250 GeV for 
     the standard model case. The pions are the analogs of the longitudinal 
     parts of the W and Z mesons while the sigma is the analog of the Higgs
     boson. For this analogy to be most direct, one would expect the 
     Higgs sector of the standard model not to be a fundamental entity but
     to represent an effective description of a technicolor model
     \cite{tm} of some sort.
     In addition, the most common treatment of the SU(2) linear sigma
     model for pion physics considers the (non-linear) limit of the model, 
     where the sigma mass is sent to infinity, to be the correct one.

         In any event, the status of the light scalars in QCD is
          quite interesting
     in itself. Perhaps it is best to wait for results to emerge 
     from LHC (or Fermilab)
     and let experiment tell us whether this close analogy \cite{abdou}
      exists.

         Returning to pi pi scattering, consider an approach using
      the more conventional  non-linear sigma model and compute
      the real part, $R_0^0$  of the I=0, J=0 amplitude. This is shown
      as the solid line in figure 1
                   of \cite{hss}.  
      Very close to the threshold at 280 MeV, one gets the 
      good "current algebra" result. However the curve runs away fast
      and starts to violate the unitarity bound, $R_0^0<1/2$ already
      at about 500 Mev. Adding the effect of the $\rho$(770) meson
      (dashed line) is seen to be in the right direction to restore
      unitarity but way too small to succeed in the low energy region.
      There seems to be no way to save unitarity but to include the 
      effect of a light sigma resonance in the 550 MeV region 
      as shown, together with the data,
      in figure 2 of that paper. Notice that in the non-linear model framwork
      it is still possible to add the sigma in a consistent way. Finally
      notice, from figure 4 of that paper, that the addition of the accepted
       $f_0$(980) resonance 
      results in convincing agreement with experiment over the large range: 
      threshold - 1100 MeV.
      
          A similar treatment of $\pi$ - K scattering in the non-linear SU(3)
          sigma model \cite{BFSS1} yielded evidence for a strange analog of
          the sigma, the kappa. Furthermore, adding the well established scalar,
          a$_0$(980) yields \cite{BFSS2} a putative full nonet of light
          scalar mesons. But it is a somewhat puzzling one.
   
   The scalar puzzle is
 the unusual spectroscopy of the light scalar nonet.
 At present, the scalars below 1 GeV
appear to fit into a nonet as:
\begin{eqnarray}
I=0: m[f_0(600)]&\approx& 500\,\,{\rm MeV}
\nonumber \\
I=1/2:\hskip .7cm m[\kappa]&\approx& 800 \,\,{\rm MeV}
\nonumber \\
I=0: m[f_0(980)]&\approx& 980 \,\,{\rm MeV}
\nonumber \\
I=1: m[a_0(980)]&\approx& 980 \,\,{\rm MeV}
\label{scalarnonet}
\end{eqnarray}
This level ordering is seen to be flipped
 compared to that
 of the standard vector
meson nonet:
\begin{eqnarray}
I=1: m[\rho(776)]&\approx& 776\,\,{\rm MeV}\hskip .7cm
n{\bar n}
\nonumber \\
I=0: m[\omega(783)]&\approx& 783 \,\,{\rm MeV}\hskip .7cm
n{\bar n}
\nonumber \\
I=1/2: m[K^*(892)]&\approx& 892 \,\,{\rm MeV}\hskip .7cm
n{\bar s}
\nonumber \\
I=0: m[\phi(1020)]&\approx& 1020 \,\,{\rm MeV}\hskip .7cm
s{\bar s}
\label{vectornonet}
\end{eqnarray}
Here the standard quark content (n stands for a
 non-strange quark while s stands for a strange
quark) is displayed at the end for each case. The
vector mass ordering is seen to just correspond
 to the number of s-type quarks in each state. 
 It was pointed out a long time ago in Ref. \cite{j},
that the level order is automatically flipped
when mesons are made of two quarks and two antiquarks
instead of a single quark and antiquark.
Note
that, in the four quark picture, the states in
 Eq.(\ref{scalarnonet}) consecutively have
 the quark contents: $nn{\bar n}{\bar n}$,
 $nn{\bar n}{\bar s}$, $nn{\bar s}{\bar s}$
and $nn{\bar s}{\bar s}$.

In order to confirm our calculations using the
    non-linear sigma model we redid them 
    \cite{BFMNS01} using the linear
    SU(3) sigma model. In order to unitarize the resulting 
    tree level amplitudes we used the K-matrix approach
    which, from the standpoint of believability has the nice
    feature that it does not introduce
    any additional parameters. An equivalent method of unitarization
     had been previously used \cite{as} in the SU(2) linear sigma
     model.

     As another part of the puzzle one notes that
the masses of the putative scalar nonet members are
 significantly
lower than the other (tensor and two axial vector)
p-wave quark-antiquark nonets. There are enough other
scalar candidates [a$_0$(1450), K$_0$(1430) and two of 
f$_0$(1370), f$_0$(1500,) f$_0$(1710)] to make another
nonet although the masses of its
contents seem somewhat higher than an 
expected
scalar p-wave nonet. Based on the usual effect that
two mixing levels repel as well
as some more detailed features, it was suggested 
\cite{BFS3} that a global picture of these scalars
might consist of a lighter ``four quark" nonet mixing
with a heavier ``two quark" nonet.

     A field theoretic toy model to study these features
was introduced in \cite{sectionV}. This model uses a generalized
SU(3) linear sigma model and involves two different nonets: 
$M$ describes both pseudoscalars and scalars containing two
quarks while $M'$ describes both pseudoscalars and scalars
containing four quarks. There is the interesting feature
that the 2 quark vs 4 quark content of each particle is a 
prediction.
Further work
in this direction has also been presented by a
 number of authors \cite{mixing}-\cite{thermo}.

   The $M$-$M'$ model is a complicated one so we studied it
   at different levels of approximation \cite{manyfjs}.
    Also there turned out to be an
   interesting connection to instanton physics \cite{inst}. 
As a brief summary
it may be desirable to 
 just display
the  ``typical" results of \cite{frs09}. These are the 
masses and the ``two quark" vs. 
``four quark" percentages of the members of
all four nonets (light and heavy pseudoscalars
and light and heavy scalars). They are listed
in Tables \ref{phi_content_1215}  and  
\ref{s_content_1215}. Isospin but not SU(3)
symmetry is being assumed. Note that for the 
I=1/2 and I=1 states, the prime denotes the heavier
particle. For the I=0 particles there are four
states of each parity and they are denoted 
by subscripts 1, 2, 3, 4 in order of increasing 
mass.
Altogether, considering the isospin degeneracy,
there are 16 different masses. The 8 inputs
comprise the pion decay constant, the four masses:
[$m_\pi$, $m_{\pi'}$, $m_a$, $m_a'$ ], the 
strange to
non strange quark mass ratio (which is related to
assuming a value for $m_K$) and the sum and the product of all
the four I=0 pseudoscalar squared masses (Each possible scenario
for their identification with physical states was considered).

    It is encouraging that the predictions seem
    to be smooth continuations of those obtained in
    earlier simplified analyses containing just zero quark mass
     terms and SU(3) degenerate non-zero quark mass terms.
     Comparing the pseudoscalar $\pi$-$\pi'$ system
     with the scalar "partner" $a$-$a'$ system, for example, one
     sees that the low mass pion is predominantly of 2 quark nature
     while the low mass a meson is predominantly of four
     quark nature. The situation is reversed for the higher mass
     states $\pi'$ and $a'$. It is the same story if one compares the 
     $K$-$K'$ system with the scalar $\kappa$-$\kappa'$ system.
     The lightest of the four mixing scalar singlets, the $f_1$
      is to be identified with the "sigma". Actually, the mass 
      listed is a "bare" one. Unitarization of the pi-pi 
      scattering amplitude gives a complex pole position, 
      $z=M^2-iM\Gamma$ with M = 477 MeV and $\Gamma$ = 504 MeV,
      which is roughly like the value extracted from the experimental data
      in \cite{cgl}.

        While it appears a little unusual to think of, say, the ordinary 
        pion as having some four quark content when treated in an 
        effective Lagrangian framework, that is in fact the standard 
        picture in the parton model approach to QCD. In the 
        case of the two scalar nonets, the mass ordering itself
         naturally suggested such a picture. This picture 
         was then inherited by the pseudoscalars when we chose to
         describe the scalars via a linear sigma model. 

        We would like to note that the subject of light scalar 
        meson spectroscopy has received a lot of attention in the 
        last 15 years and that a more complete
         documentation of this  recent work is given in \cite{fpss}.

\begin{table}
	\begin{tabular}{c|c|c|c}
		\hline \hline
		State & \,\, ${\bar q} q$\% \, \, & \, ${\bar q} 
		{\bar q} q q$\%\, & \, $m$ (GeV)
		\\
		\hline
		\hline
		$\pi$      & 85  & 15 & 0.137\\
		\hline
		$\pi'$      & 15  & 85 & 1.215\\
		\hline
		$K$          & 86  & 14 & 0.515\\
		\hline
		$K'$          & 14  & 86 & 1.195\\
		\hline
		$\eta_1$   &  89   &  11   & 0.553 \\
		\hline
		$\eta_2$   &  78   &  22   & 0.982\\
		\hline
		$\eta_3$   &  32   &  68   &  1.225 \\
		\hline
		$\eta_4$   &  1   & 99     & 1.794\\
		\hline
		\hline
	\end{tabular}
	\caption{
	Predicted properties of pseudoscalar
	states:
	${\bar q} q$ percentage (2nd
	column), ${\bar q} {\bar q} q q$ (3rd
	column) and masses (last column).}
	\label{phi_content_1215}
\end{table}

\begin{table}
	\begin{tabular}{c|c|c|c}
		\hline \hline
		State &\,\, ${\bar q} q$\%\,\,& \, ${\bar q} 
		{\bar q} q q$\% \, & \, $m$ (GeV)
		\\
		\hline
		\hline
		$a$        &  24 &  76 & 0.984   \\
		\hline
		$a'$        &  76 &  24 & 1.474   \\
		\hline
		$\kappa$  &   8  &  92  & 1.067  \\
		\hline
		$\kappa'$  &   92  &  8  & 1.624 \\
		\hline
		$f_1$     &  40    &  60  & 0.742  \\
		\hline
		$f_2$     &  5   &    95  & 1.085 \\
		\hline
		$f_3$     &  63   &   37  & 1.493 \\
		\hline
		$f_4$     &  93   &   7   & 1.783 \\
		\hline
		\hline
	\end{tabular}
	\caption{Predicted properties of scalar 
	states: 
	${\bar q} q$ percentage (2nd 
	column), ${\bar q} {\bar q} q q$ (3rd 
	column) and masses (last column).
	 }
	\label{s_content_1215}
\end{table}

\begin{theacknowledgments}
We owe a great deal of thanks to our collaborators on various parts of
this work:
A. Abdel-Rehim, D.Black, M. Harada, S. Moussa, S. Nasri,
N.W. Park, F. Sannino and M.N.Shahid.

J.S. would like to thank the organizers for
 inviting him to 
this very valuable and pleasant workshop and
 is grateful to the Academia Mexicana
de Ciencias for the financial support to
participate.
 The work was supported in part by 
 the US DOE under Contract No. DE-FG-02-85ER 40231.
 J.S. is also happy 
to thank the CP3- Origins group at the University
of Southern Denmark for their warm hospitality and 
partial support during the Fall 2009 
academic semester.
\end{theacknowledgments}

\bibliographystyle{aipproc}

\end{document}